\documentclass[11pt,twoside]{article}
\usepackage{./asp2010}

\resetcounters

\begin{document}

\title{A couple of recent developments in the structure of the outer disk of the Milky Way}
\author{Giovanni Carraro$^1$, Gabriel Perren$^2$, Rub\'en A. V\'azquez$^2$, Andre Moitinho$^3$ 
\affil{$^1$ESO, Alonso de Cordova 3107, 19001, Santiago de Chile, Chile\\
$^2$Facultad de Ciencias Astron\'omicas y Geof\'isicas (UNLP), Instituto de Astrof\'isica de La Plata (CONICET, UNLP), Paseo del Bosque s/n, La Plata, Argentina\\
$^3$SIM/IDL, Faculdade de Ci\^{e}ncias da Universidade de Lisboa, Ed. C8, Campo Grande, 1749-016 Lisboa, Portugal}}

\begin{abstract}
In this contribution we  summarize recent achievements by our group on the understanding of the structure of the outer Galactic disk, with particular emphasis to the outer disk {\it extent}, and the spiral
structure beyond the solar circle.
\end{abstract}

\section{Spiral structure in the outer Galactic disk}

The spiral structure of the Milky Way beyond the solar ring has traditionally been over-looked. HI surveys mostly concentrated in the inner disk, where the HI density is larger and detections easier.
Despite many years of  accumulating data, the debate on how many arms the Galaxy has, and their nature, shape and pitch angles, is still lively \citep{Turner2013}.\\
 
\begin{figure}[!ht]
\plotone{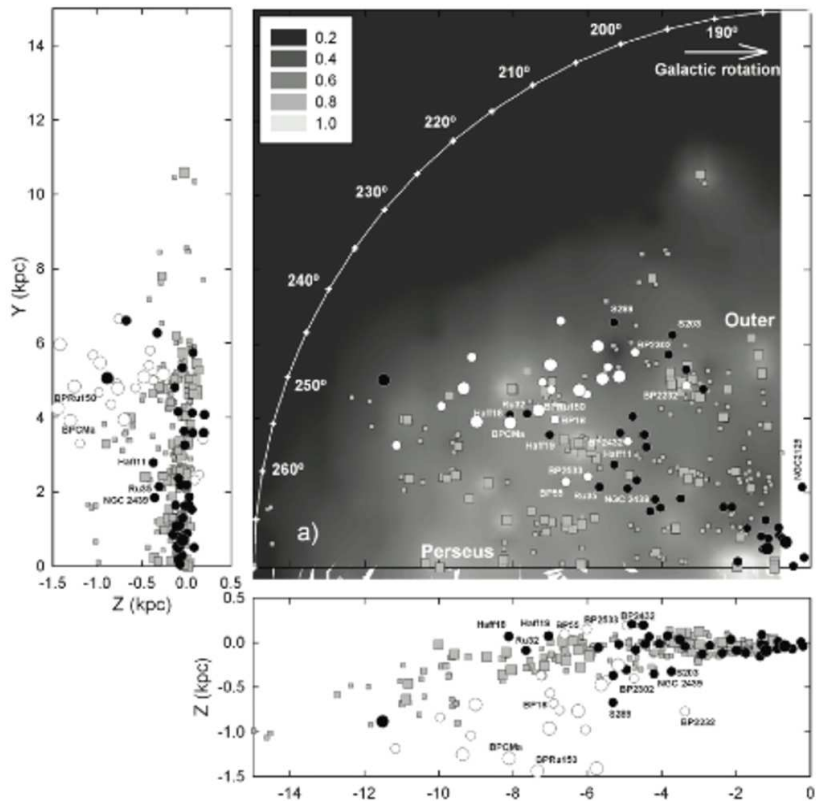}
\caption{The spiral structure of the Galactic disk in the third quadrant from \citet{Vazquez2008}. Solid dots indicate open clusters, while white dots
 indicate associations of young stars in the background of open star clusters. Finally, squares are for CO clouds. Notice the clear indication of the thin disk warp in the bottom panel.}
\end{figure}

\noindent
The situation in the outer disk improved significantly in the last 10 years, especially in the southern Milky Way. 
One of the most remarkable results is the  detection of the fourth quadrant distant arm by \citet{McClure-Griffiths2004} in HI. 
A whole sky HI map was presented by \citet*{Levine2006} which covers also the outer disk. \citet{Vazquez2008} presented additional radio data using the CO molecules in the outer disk, in the third quadrant. 
Together with CO, \citet{Vazquez2008} also presented optical data from a large sample
of young open clusters and young associations located behind these clusters.
This wealth of data allows one to describe the outer disk spiral structure in some details. Interestingly enough, the outcome is  different from the GLIMPSE description of the MW spiral arm, popularized by 
\citet[their Fig 15]{Churchwell2009}.\\

This study is based on counting Red Giant clump stars all over the Milky Way.
Red Giant clump stars span ages from 0.5 to 10 Gyrs, so they virtually trace several different stellar populations. 
If they do trace spiral structure, this would imply that spiral arms are long-lived structure, something that simulations do not confirm.
\citet{Carraro2013} discuss in details the nature of these stars, concluding that they are not ideal spiral m tracers.
Looking in fact at external galaxies, spiral arms possess plenty of gas and blue, hence extremely young, stars.
According to \citet{Churchwell2009} the Galaxy has two majors arms,  Perseus and Scutum Crux, and two additional minor arms, Carina-Sagittarius and Norma Cygnus.
The conclusion is then that the Milky Way is a two-arms spiral. \\

\noindent
At odds with \citet{Churchwell2009}, surveys like those of \citet{Levine2006} and \citet{Vazquez2008} show that Sagittarius is more prominent than Scutum Crux, and that Perseus is negligible compared to Norma Cygnus.
This confirms the tradition that tracing spiral arms in our own Galaxy is challenging and complicated. And that the outcome still depends on the adopted tracers.\\

\noindent
However, if we have to believe to what we see outside the Milky Way, the evidences from optical and radio surveys (HI, HII, CO) should be preferred.
The picture emerging from these surveys  (see Fig ~1  and \citet{Levine2006} Fig.~4A,B) is that the Norma-Cygnus arm (also called outer arm) is prominent in the third quadrant, and its ideal prolongation 
encounters the distant arm detected by \citet{McClure-Griffiths2004}  in
the fourth quadrant.  Besides, there are very minor indications of a Perseus arm in the third quadrant, and virtually none in the fourth.
What is prominent in the third quadrant is a feature which resembles an extension of the local arm, visible both in optical \citep{Vazquez2008} and in HI \citep{Levine2006}.
The nature of this structure is unclear. Available observations show that it is most probably an inter-arm feature, a bridge, and not a proper arm. Its origin is not understood, but Purcell during the conference showed that it can have a tidal origin.

The spiral structure in the third quadrant is referred to as an excess of star formation (Benjamin, this conference) that artist's conception cannot reproduce. 
We believe this is not the case, but, simpler, it is so because we are looking at a privileged  direction,
where extinction has been traditionally known to be low \citep{Fitzgerald1968, Janes1991, Moitinho2001}, thus permitting to see very far from the Sun.

\section{The outer disk: density break, cut-off, truncation?}

During the conference several speakers mentioned the fact that almost all disk galaxies show density breaks in the form of a  change of slope in the disk surface brightness  profile.
According to GLIMPSE data \citep{Churchwell2009, Benjamin2013} this occurs also in our Milky Way, at a distance of about 13-14 kpc from the Galactic center.
This distance coincides with the cut-off radius of the disk adopted in Galactic models like Besan\c{c}on \citep{Robin1992}, and is referred to as the edge of the stellar disk \citep{Minniti2011}.

Star clusters, associations and gas are routinely found well beyond this distance \citep{Carraro2010, Zasowski2013}, and therefore  naming this distance as {\it edge}
or {\it truncation} of the disk is highly inappropriate. The cut-off, edge or truncation used in Galactic models is caused by not taking into account the disk warp and flare \citep{Lopez-Corredoira2012},
while claims like the ones in \citet{Minniti2011} are simply based on the incapability to detect correctly red clump stars below some magnitude limit, due to confusion and photometric errors.\\

\noindent
The actual difficulties of Galactic models to reproduce star counts in the outer disk is shown in Fig~2 \citep{Perren2013}, where star counts is several Galactic anti-center directions are compared with predictions from Besan\c{c}on and TRILEGAL (Girardi et al. 2005) models. It is evident how much work is still needed on the model side.

\begin{figure}[!ht]
\plotone{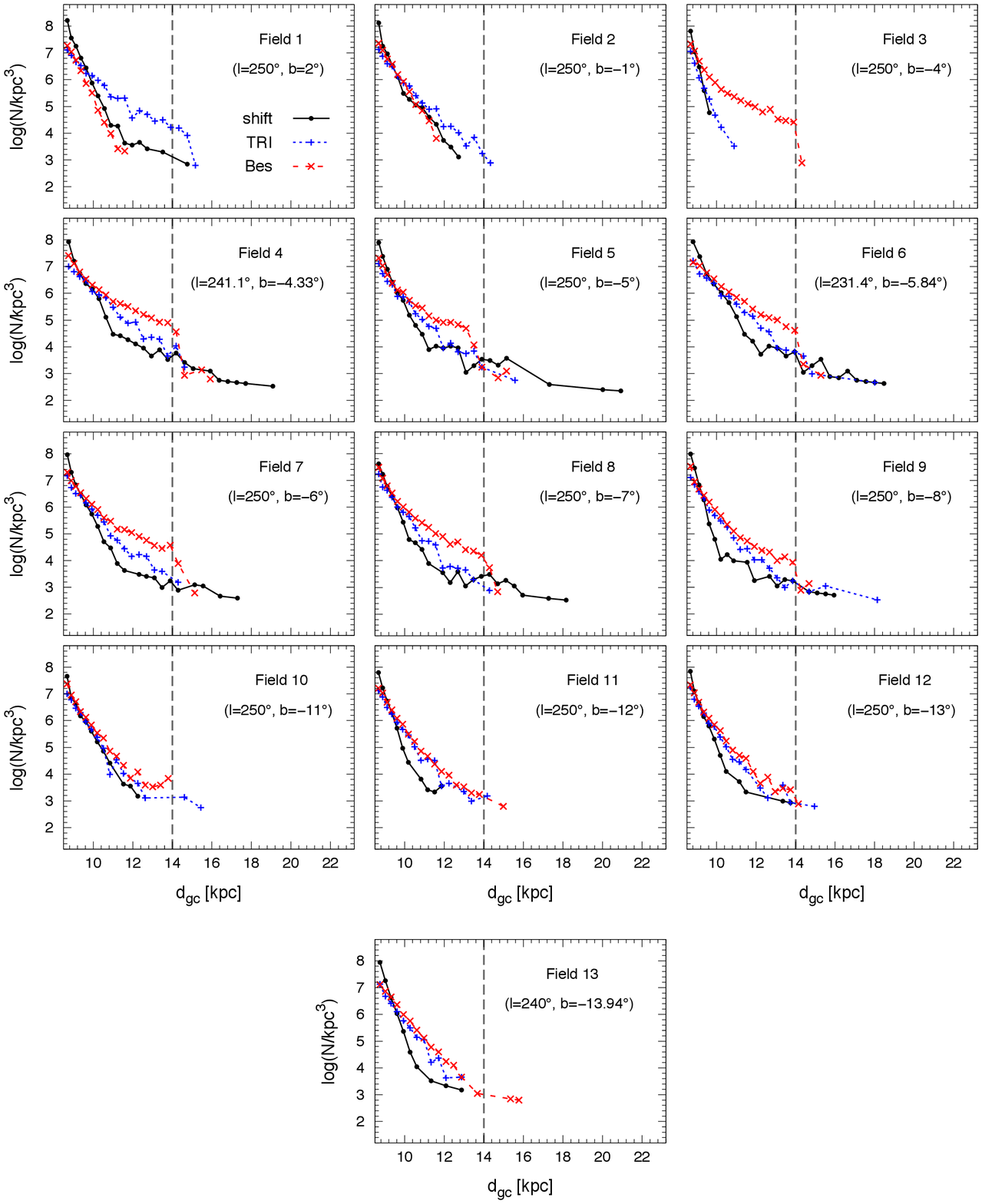}
\caption{Star counts along several directions in the third Galactic quadrant (solid black lines). Red and blue lines indicate the expectation from Besancon and Trilegal models, respectively.}
\end{figure}

\acknowledgements  G. Carraro acknowledges useful discussions with V. Korchagin, B. Burton and R. Benjamin.

\bibliography{gcarraro}

\end{document}